\documentclass[10pt,a4paper,onecolumn]{article}

\usepackage[utf8]{inputenc}
\usepackage[T1]{fontenc}
\usepackage{lmodern}
\usepackage{amsmath,amssymb}
\usepackage{graphicx}
\usepackage[rgb,svgnames]{xcolor}
\usepackage[authoryear]{natbib}
\usepackage{authblk}
\usepackage{titlesec}
\usepackage{caption}
\usepackage{float}
\usepackage{hyperref}
\usepackage{tikz}

\hypersetup{
    unicode=true,
    pdftitle={pyKurucz: A Pure Python Reimplementation of Kurucz ATLAS12 and SYNTHE for Self-Consistent Stellar Atmospheres and Spectra},
    pdfauthor={Elliot M. Kim, Yuan-Sen Ting},
    colorlinks=true,
    linkcolor=Maroon,
    citecolor=Blue,
    urlcolor=Blue,
    breaklinks=true
}

\usepackage[top=2.5cm, bottom=2.5cm, left=2.5cm, right=2.5cm]{geometry}

\titleformat{\section}
  {\normalfont\sffamily\Large\bfseries}{}{0pt}{}
\titleformat{\subsection}
  {\normalfont\sffamily\large\bfseries}{}{0pt}{}
\titleformat{\subsubsection}
  {\normalfont\sffamily\bfseries}{}{0pt}{}
\titleformat*{\paragraph}{\sffamily\normalsize}

\let\origfigure\figure
\let\endorigfigure\endfigure
\renewenvironment{figure}[1][2]{%
    \expandafter\origfigure\expandafter[H]
}{%
    \endorigfigure
}

\usepackage{parskip}

\definecolor{c53baa1}{RGB}{83,186,161}
\definecolor{c202826}{RGB}{32,40,38}
\def\rorglobalscale{0.1}
\newcommand{\rorlogo}{%
\begin{tikzpicture}[y=1cm, x=1cm, yscale=\rorglobalscale,xscale=\rorglobalscale, every node/.append style={scale=\rorglobalscale}, inner sep=0pt, outer sep=0pt]
  \begin{scope}[even odd rule,line join=round,miter limit=2.0,shift={(-0.025, 0.0216)}]
    \path[fill=c53baa1,nonzero rule] (1.8164, 3.012) -- (1.4954, 2.5204) -- (1.1742, 3.012) -- cycle;
    \path[fill=c53baa1,nonzero rule] (3.1594, 3.012) -- (2.8385, 2.5204) -- (2.5172, 3.012) -- cycle;
    \path[fill=c53baa1,nonzero rule] (1.1742, 0.0669) -- (1.4954, 0.5588) -- (1.8164, 0.0669) -- cycle;
    \path[fill=c53baa1,nonzero rule] (2.5172, 0.0669) -- (2.8385, 0.5588) -- (3.1594, 0.0669) -- cycle;
  \end{scope}
\end{tikzpicture}%
}

\title{\sffamily\bfseries pyKurucz: A Pure Python Reimplementation of Kurucz ATLAS12 and SYNTHE for Self-Consistent Stellar Atmospheres and Spectra}

\author[1]{Elliot M.\ Kim}
\author[2,3]{Yuan-Sen Ting}
\affil[1]{Cornell University, USA\,\protect\href{https://ror.org/05bnh6r87}{\protect\rorlogo}}
\affil[2]{The Ohio State University, USA\,\protect\href{https://ror.org/00rs6vg23}{\protect\rorlogo}}
\affil[3]{Max Planck Institute for Astronomy, Heidelberg, Germany\,\protect\href{https://ror.org/01vhnrs90}{\protect\rorlogo}}

\date{11 May 2026}

\begin{document}
\sloppy  
\maketitle

\section{Summary}

pyKurucz is a pure Python reimplementation of Kurucz's ATLAS12 and SYNTHE --- the standard tools for computing self-consistent stellar atmospheres and synthetic spectra from first principles. Stellar spectra encode information about a star's temperature, gravity, and chemical composition, but extracting it requires a forward model: starting from $(T_{\rm eff}, \log g)$ and an abundance pattern, the temperature--pressure--density stratification of the atmosphere must first be solved (radiative/convective equilibrium with opacity sampling), and then the emergent spectrum must be computed by integrating radiative transfer through that atmosphere wavelength by wavelength. The Kurucz Fortran codes have filled this role for decades. pyKurucz performs the same calculation, to the same numerical precision, entirely in Python.

The package consists of two cooperating engines. \textbf{\texttt{atlas\_py}} is the atmosphere engine: it iteratively solves Saha--Boltzmann populations, molecular equilibrium, continuous opacity (H$^-$, H\,\textsc{i} bound-free via Karzas \& Latter cross-sections \citep{karzas1961}, He, metals, hydrogen and helium Rayleigh, H$_2$ Rayleigh, Thomson), atomic line opacity by \emph{direct opacity sampling} from preselected Kurucz line records (LINOP/XLINOP), the Rosseland mean, the radiative transfer equation (JOSH solver, using the same pretabulated BLOCKJ/BLOCKH coefficient matrices as ATLAS12), and mixing-length convection, then applies temperature corrections (TCORR) until the layer structure converges. Direct opacity sampling --- the defining feature of ATLAS12 over the more common ATLAS9 --- means the atmosphere is recomputed against the \emph{actual} abundance pattern requested, so peculiar patterns (CEMP, Ap, individual $\alpha$-element offsets, r-process enhancements) are handled identically to bulk-scaled ones, with no opacity-distribution-function (ODF) re-tabulation required. \textbf{\texttt{synthe\_py}} is the synthesis engine: given a converged atmosphere it evaluates continuous opacity, line opacity from the full Kurucz GFALL atomic line list with Voigt profiles (thermal $+$ van der Waals $+$ Stark $+$ radiative broadening), Stark-broadened hydrogen line profiles via the HPROF4 routine, tabulated helium line profiles (BCS, Griem, and Dimitrijevi\'c lookup tables), molecular line opacity from the Schwenke TiO list, the Partridge--Schwenke H$_2$O list, and the GFALL diatomic pack (CN, CO, C$_2$, CH, OH, MgH, FeH, and others), and the radiative transfer equation on a high-resolution wavelength grid via its own JOSH solver. The end-to-end orchestrator (\texttt{pykurucz.py}) chains an optional neural warm-start \citep[kurucz-a1;][]{li2025} into \texttt{atlas\_py} into \texttt{synthe\_py} to go from $(T_{\rm eff}, \log g, \rm[M/H], [\alpha/M]$, per-element abundances$)$ to a synthetic spectrum.

The implementation uses NumPy, SciPy, and Numba (JIT compilation of inner loops in both engines), requiring no Fortran compiler. End-to-end, on five representative test cases spanning 2500\,K cool giants to 44000\,K O stars, the Python pipeline reproduces Fortran ATLAS12 $+$ SYNTHE to sub-0.005\% median deviation in normalized flux over 300--1800\,nm at resolving power $R = 300{,}000$. Full documentation --- installation, user guide, architecture, physics reference, and worked examples --- is available at \url{https://pykurucz.vercel.app}.

\section{Statement of need}

Kurucz's ATLAS12 (atmospheres) and SYNTHE (synthesis), together with the atomic and molecular line lists that accompany them, are the de facto standards for stellar-spectrum forward modeling, accumulating tens of thousands of citations over several decades \citep{kurucz2005}. The original Fortran, written in a pre-Fortran-77 dialect with fixed-format source, single-character variables, computed GOTOs, and EQUIVALENCE statements, is incompatible with modern \texttt{gfortran} without significant manual patching --- compiling it is a rite of passage that has defeated many research groups. Following the passing of Robert L.\ Kurucz in 2025, the long-term maintenance trajectory of the original code is uncertain. This work provides a faithful, ground-truth numerical reimplementation of \emph{both} ATLAS12 and SYNTHE in pure Python --- not a wrapper around Fortran, but a line-by-line translation validated against the original.

A crucial property of the package is \textbf{self-consistency between abundances and atmospheric structure}. Because abundances enter the opacity, they reshape the temperature--pressure stratification, which in turn rescales every line in the spectrum --- not just the lines of the perturbed elements. Treating the atmosphere as fixed while fitting abundances breaks this loop. ATLAS12's direct opacity sampling closes it: pyKurucz exposes both bulk scaling (\texttt{--mh}/\texttt{--am}) and arbitrary per-element overrides (\texttt{--abund Fe:-1.0 --abund C:+0.4}) through the same \texttt{atlas\_py} iteration, so the atmosphere always matches the abundance pattern used by the synthesis. This is the property that distinguishes ATLAS12 from the more widely distributed ATLAS9, and the reason pyKurucz reimplements ATLAS12 specifically.

A Python reimplementation offers concrete benefits over Fortran: direct integration with machine learning workflows (training emulators, differentiable spectral fitting); interoperability with the broader Python ecosystem (NumPy, SciPy, Astropy); flexibility for customization without recompiling Fortran; cross-platform installation with no compiler dependency; and improved readability for teaching and debugging.

\section{State of the field}

Stellar atmosphere modeling and spectrum synthesis have long been dominated by Fortran-based codes. On the atmosphere side, the most widely used grids and engines are Kurucz's ATLAS9/ATLAS12 \citep{kurucz2005}, MARCS, and PHOENIX. On the synthesis side, the standards are SYNTHE \citep{kurucz2005}, MOOG \citep{sneden1973}, Turbospectrum \citep{plez2012}, and SPECTRUM \citep{gray1999}. While highly accurate, these codes require Fortran toolchains, are difficult to install across operating systems, often lack documentation, and do not integrate naturally with modern Python-based analysis workflows.

Several efforts have brought synthesis (but not the underlying atmosphere) into more modern environments. Spectroscopy Made Easy \citep[SME;][]{piskunov2017} combines an IDL interface with compiled C++ and Fortran backends; its Python rewrite, PySME \citep{wehrhahn2023}, replaced the IDL layer but retained the compiled synthesis engine. iSpec \citep{blancocuaresma2014} provides a Python framework wrapping SYNTHE, MOOG, or Turbospectrum, but requires the underlying Fortran codes to be installed. KORG \citep{wheeler2023} is a modern 1D LTE synthesis package written in Julia with automatic differentiation support; it represents the closest analog in philosophy, though it is a synthesis-only code that consumes pre-computed atmospheres and uses a different language ecosystem. None of these tools include a self-consistent atmosphere solver --- they assume the atmosphere is provided externally.

A separate class of approaches uses machine learning to approximate synthetic spectra. The Cannon \citep{ness2015} learns a data-driven spectral model from a training set; The Payne \citep{ting2019} trains a neural network to interpolate ab initio models across label space. These are powerful for survey analysis but are statistical approximations tied to training-grid fidelity. The kurucz-a1 emulator \citep{li2025} used as an optional warm-start in pyKurucz belongs to this class on the atmosphere side: it predicts an ATLAS12-quality starting structure from $(T_{\rm eff}, \log g, \rm[M/H], [\alpha/M])$ to accelerate convergence, but the actual atmosphere is then iterated by \texttt{atlas\_py} itself, so the final pipeline output is fully ab initio.

pyKurucz occupies a distinct position: it is a faithful numerical reimplementation of \emph{both} ATLAS12 and SYNTHE --- the specific codes whose line lists and physics have been the de facto standard for decades --- validated against the Fortran ground truth, requiring no Fortran at any stage. It complements the other tools above by exposing self-consistent atmosphere modelling and synthesis within a pure Python environment.

\section{Software design}

pyKurucz is organized as a modular pipeline with two cooperating engines and an optional neural warm-start.

\texttt{atlas\_py} (the atmosphere engine) implements the full ATLAS12 iteration loop. Each outer iteration follows the same eight-step sequence as the Fortran code: POPSALL (Saha--Boltzmann populations $+$ partition functions, with NMOLEC for the molecular equilibrium when MOLECULES ON), KAPCONT (continuous opacity), LINOP/XLINOP (atomic line opacity from preselected Kurucz \texttt{fort.12} records and NLTE \texttt{fort.19} lines), ROSS (Rosseland mean opacity and optical depth scale), RADIAP (radiative flux and acceleration), TCORR (temperature corrections from flux errors), CONVEC (mixing-length convection where unstable), and JOSH (frequency-by-frequency radiative transfer). HLINOP supplies hydrogen-line wing opacity. Convergence is tested on the maximum normalized change across the physical atmosphere columns (RHOX, T, P, XNE, ABROSS, VTURB) and the loop exits early once a user-supplied tolerance is met for a configurable number of consecutive iterations. The output is a converged Kurucz-format \texttt{.atm} file.

\texttt{synthe\_py} (the synthesis engine) is organized into three layers. The \textbf{I/O layer} (\texttt{synthe\_py.io}) handles atmosphere loading (Kurucz \texttt{.atm} and preprocessed \texttt{.npz} formats), atomic line catalog parsing and compilation, and spectrum export. The \textbf{physics layer} (\texttt{synthe\_py.physics}) implements continuous opacity interpolation (KAPP), line opacity with Voigt profiles (\texttt{voigt\_jit.py} for Numba-accelerated evaluation), hydrogen Stark-broadened profiles (HPROF4/HLINOP port), helium tabulated line profiles (BCS, Griem, and Dimitrijevi\'c lookups), autoionizing line profiles, Saha--Boltzmann populations with detailed partition functions (PFSAHA), and the molecular equilibrium solver. The \textbf{engine layer} orchestrates wavelength-by-wavelength opacity accumulation and radiative transfer via the JOSH solver, yielding both the line$+$continuum flux $F_\lambda$ and the continuum-only flux $F_{\rm cont}$ (\texttt{synthe\_py.engine.radiative}).

\texttt{pykurucz.py} is the end-to-end orchestrator. It accepts stellar parameters $(T_{\rm eff}, \log g, \rm[M/H], [\alpha/M], v_{\rm turb}$, optional per-element abundances$)$, queries the kurucz-a1 neural emulator for a warm-start atmosphere, hands that to \texttt{atlas\_py} for self-consistent iteration with \texttt{MOLECULES ON}, and finally hands the converged atmosphere to \texttt{synthe\_py.cli} for the spectrum. A second entry point (\texttt{synthesize\_from\_atm.py}) bypasses the atmosphere step for users who already have a \texttt{.atm} file from any source (ATLAS12, MARCS, PHOENIX, \dots); in that mode \texttt{synthe\_py} runs standalone. The emulator is optional but strongly recommended: cold-starting \texttt{atlas\_py} from a grey approximation can stall, whereas the emulator drops the iteration into the convergence basin so it typically converges in 10--15 steps.

pyKurucz reuses the same input data as the original Fortran. The Kurucz atomic line catalogues (\textsc{gfall}, predicted lines, low- and high-excitation lists, NLTE lists), molecular line lists (Schwenke TiO, Partridge--Schwenke H$_2$O, and the GFALL diatomic pack), partition functions, ionization potentials, continuum opacity tables, and molecular equilibrium data are stored as compact binary or \texttt{.npz} archives pre-extracted from the Fortran data files, ensuring the Python pipeline operates on exactly the same physical data as the Fortran ground truth. Performance-critical routines --- Voigt profile evaluation, opacity accumulation, line selection, JOSH integration --- use Numba JIT compilation, and the wavelength-parallel radiative transfer worker pool keeps wall-time competitive with Fortran.

\begin{figure}[H]
\centering
\includegraphics[width=\textwidth]{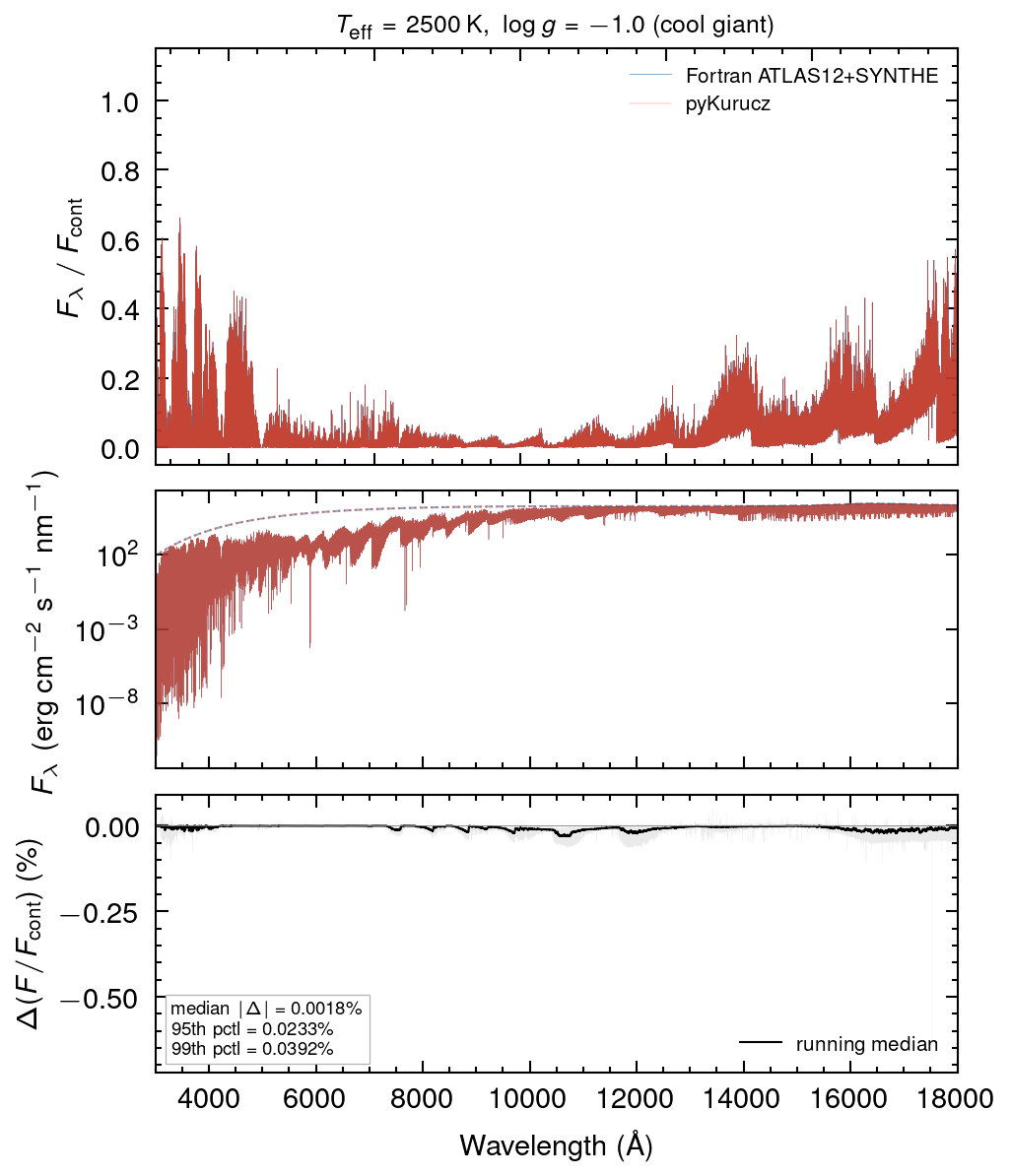}
\caption{End-to-end pyKurucz vs Fortran ATLAS12 $+$ SYNTHE for a cool giant ($T_{\text{eff}}=2500$\,K, $\log g=-1.0$). Both pipelines start from the same Kurucz grid-warmstart \texttt{.atm}, run 30 ATLAS12 outer iterations, and then synthesize 300--1800\,nm at $R = 300{,}000$. Top: normalized flux overlay (3500--9000\,\AA). Middle: full flux spectrum (3000--18000\,\AA, log scale). Bottom: fractional deviation of normalized flux with running median.}
\label{fig:cool_giant}
\end{figure}

\begin{figure}[H]
\centering
\includegraphics[width=\textwidth]{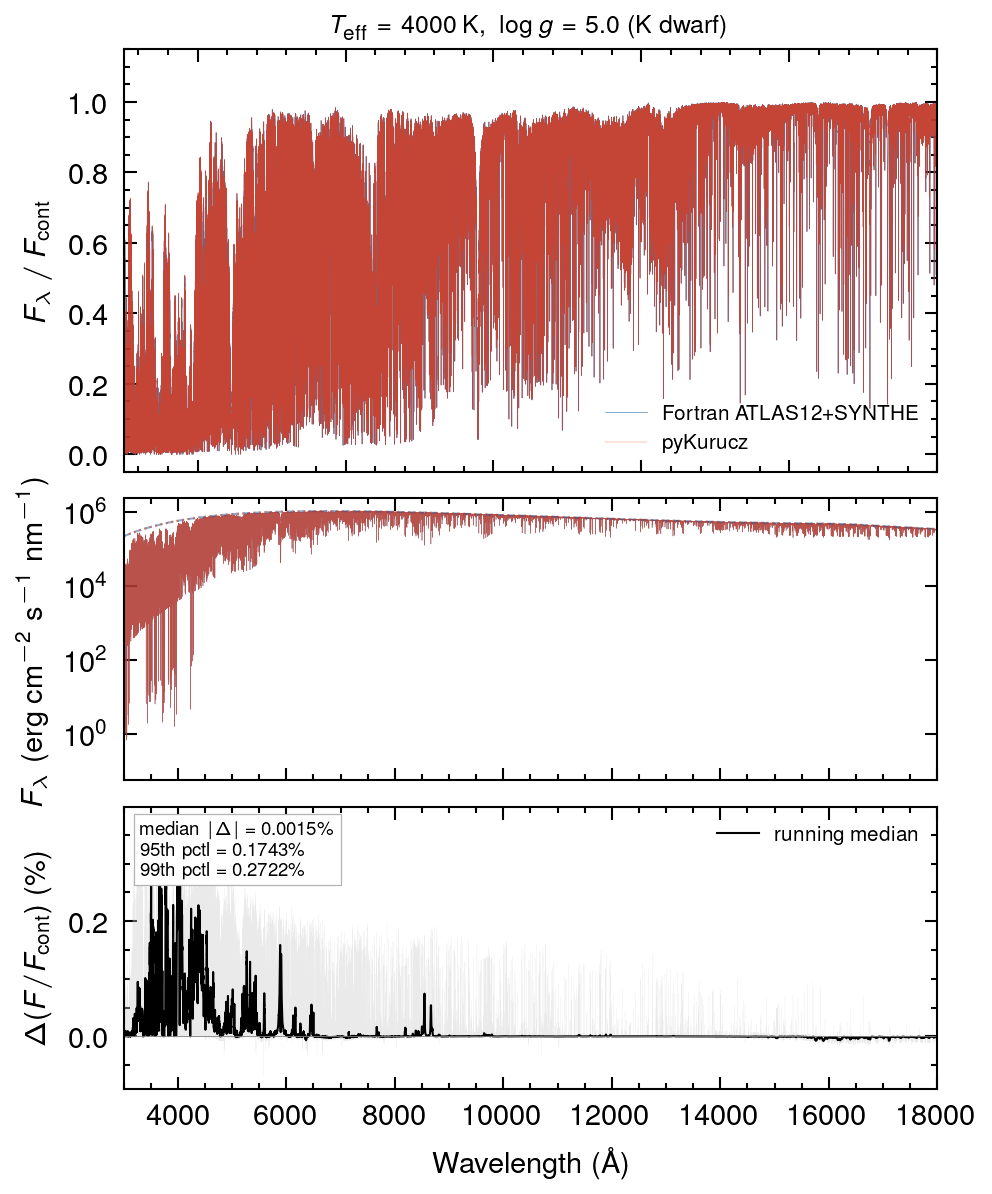}
\caption{End-to-end comparison for a K dwarf ($T_{\text{eff}}=4000$\,K, $\log g=5.00$).}
\label{fig:k_dwarf}
\end{figure}

\begin{figure}[H]
\centering
\includegraphics[width=\textwidth]{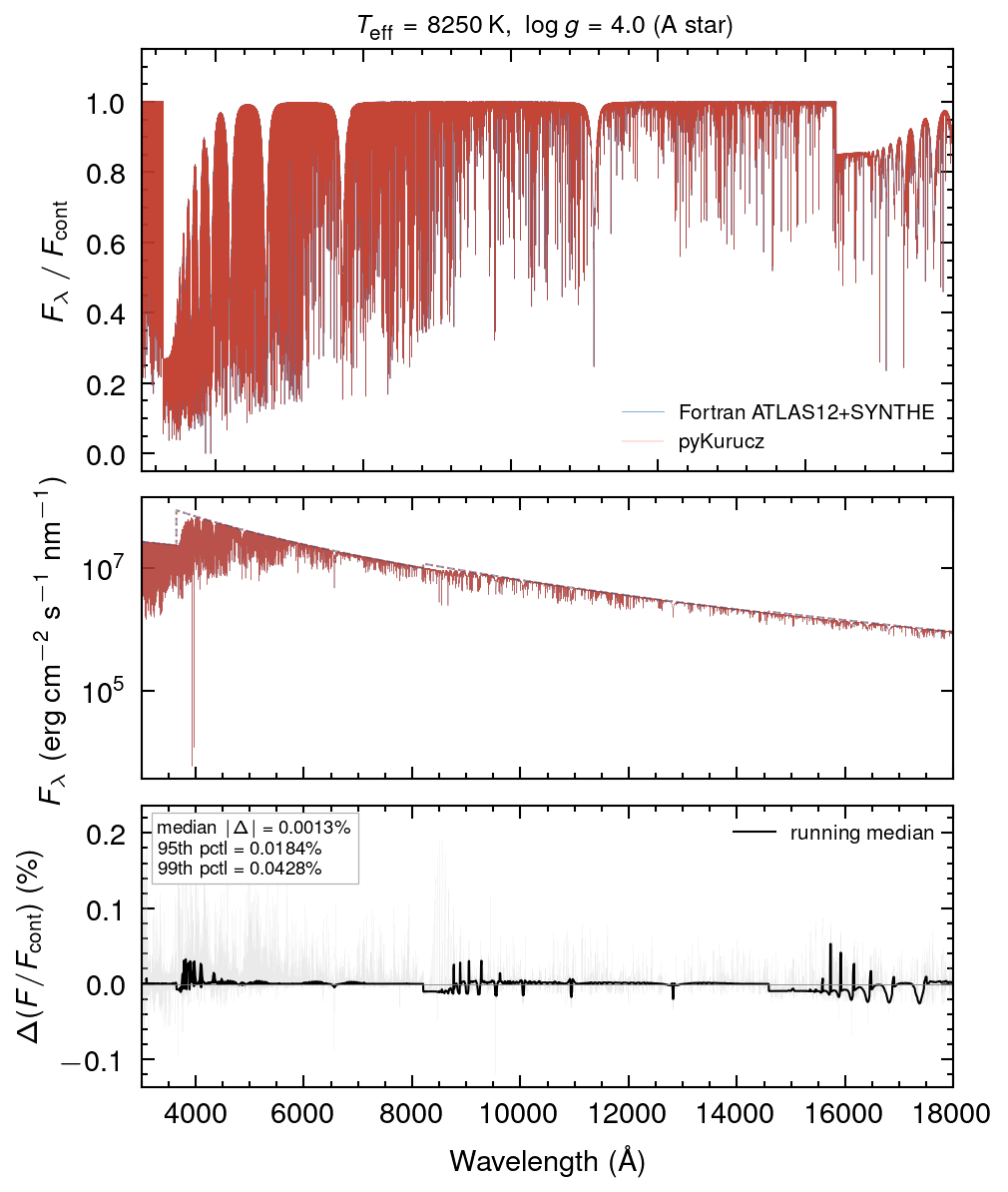}
\caption{End-to-end comparison for an A-type star ($T_{\text{eff}}=8250$\,K, $\log g=4.00$).}
\label{fig:a_star}
\end{figure}

\begin{figure}[H]
\centering
\includegraphics[width=\textwidth]{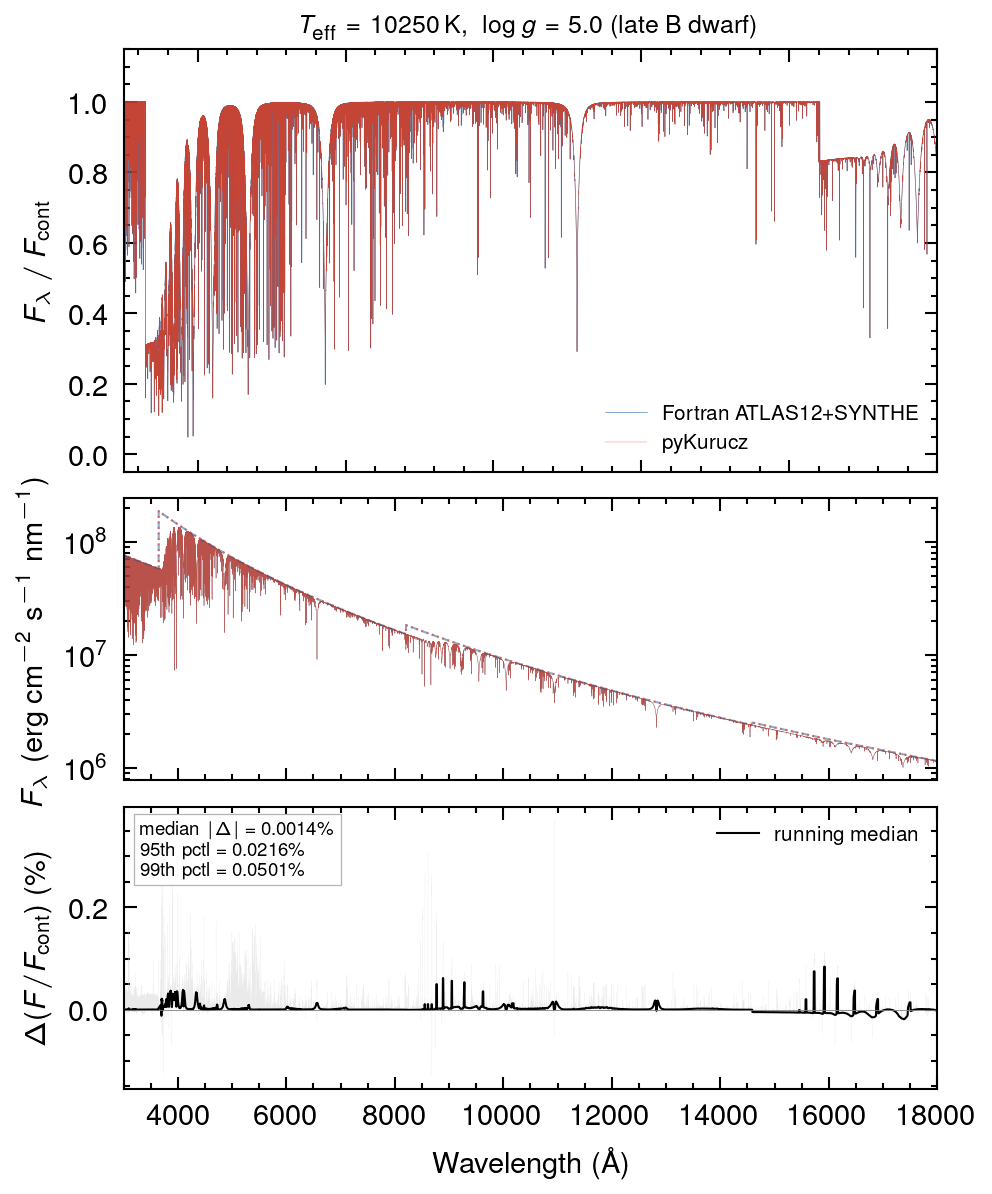}
\caption{End-to-end comparison for a late B dwarf ($T_{\text{eff}}=10250$\,K, $\log g=5.00$).}
\label{fig:late_b}
\end{figure}

\begin{figure}[H]
\centering
\includegraphics[width=\textwidth]{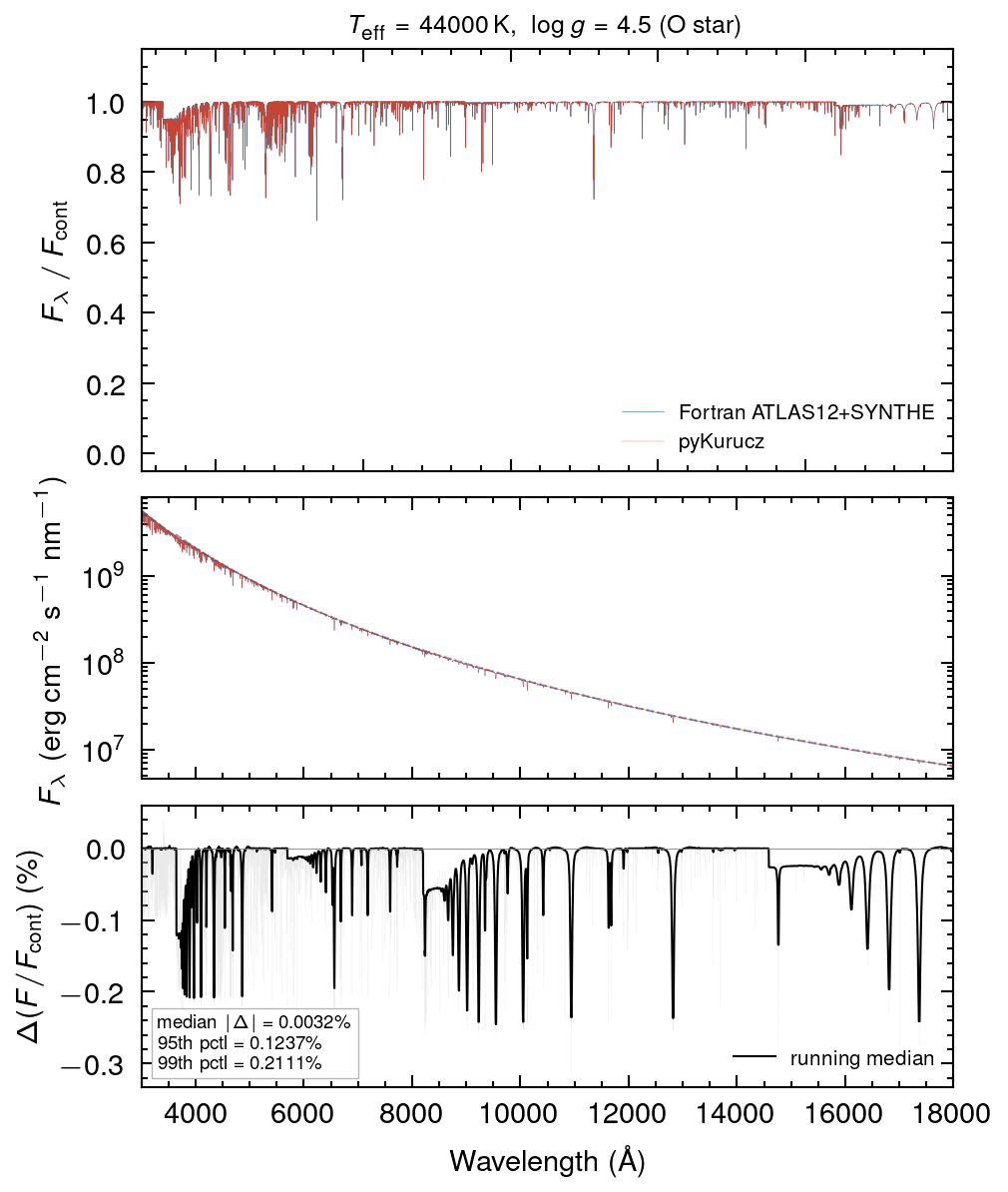}
\caption{End-to-end comparison for a hot O-type star ($T_{\text{eff}}=44000$\,K, $\log g=4.50$).}
\label{fig:o_star}
\end{figure}

\section{Research impact statement}

pyKurucz has been validated end-to-end against the Fortran ATLAS12 $+$ SYNTHE pipeline. To isolate engine-implementation differences from initial-condition differences, both pipelines are launched from the \emph{same} Kurucz grid-warmstart \texttt{.atm}, with the same microturbulent velocity, and \texttt{atlas\_py} is run for the same 30 outer iterations as the Fortran ATLAS12 deck. Five representative comparisons spanning $T_{\rm eff} = 2500$--$44000$\,K are included with the package (Figures~\ref{fig:cool_giant}--\ref{fig:o_star}). Across all five, the median normalized-flux deviation stays below 0.005\%, with 99th-percentile deviations below 0.3\% and sub-1\% maxima --- essentially the same level as the synthe-only validation against Fortran SYNTHE on a fixed atmosphere, which establishes that \texttt{atlas\_py} and \texttt{synthe\_py} reproduce Fortran ATLAS12 $+$ SYNTHE end-to-end at the level of numerical noise on these representative test cases.

The user-facing default flow in \texttt{pykurucz.py} uses the kurucz-a1 neural emulator \citep{li2025} for the warm-start, which is fast and supports arbitrary off-grid abundance patterns; the resulting spectra agree with Fortran to within a few tenths of a percent --- well below NLTE/3D/line-list systematics. Users who need bit-faithful Fortran spectra can use \texttt{synthesize\_from\_atm.py} to run \texttt{synthe\_py} on a pre-existing \texttt{.atm}, which agrees with Fortran SYNTHE at the level of numerical noise.

This opens several near-term research applications. First, pyKurucz enables generation of large \emph{self-consistent} training grids for data-driven spectral models (e.g., The Payne, The Cannon) without a Fortran toolchain, lowering the barrier for groups building emulators for surveys such as SDSS-V, DESI, 4MOST, and WEAVE --- and, crucially, allowing those grids to span peculiar abundance patterns (CEMP, Ap, r-process) that ATLAS9/ODF approaches cannot easily reach. Second, the pure Python pipeline can be embedded into gradient-based optimization frameworks (PyTorch, JAX) for differentiable spectral fitting where the atmosphere itself participates in the gradient. Third, the readable codebase facilitates classroom instruction and rapid prototyping of new physics. Finally, as a faithful preservation of ATLAS12 and SYNTHE in a modern language, pyKurucz serves as an archival reference implementation ensuring long-term reproducibility.

\section{Limitations and future work}

pyKurucz reimplements the full ATLAS12 $+$ SYNTHE path through atomic and molecular line opacity, but several limitations remain, each defining a clear direction for future development.

\paragraph{LTE only.} Both engines assume local thermodynamic equilibrium. Non-LTE effects matter for specific lines (Li\,\textsc{i} 6707\,\AA, Na D, O\,\textsc{i} triplet) in metal-poor and hot stars. Departure coefficients from external NLTE codes can be applied as corrections.

\paragraph{1D plane-parallel geometry.} Like the original Fortran, all models assume plane-parallel stratification. 3D hydrodynamic atmospheres (Stagger, CO$^5$BOLD) can be ingested by \texttt{synthe\_py} for post-processing, but \texttt{atlas\_py} itself is 1D.

\paragraph{Cool RSG $\alpha$-perturbed corner cases.} In a narrow region of parameter space --- cool ($T_{\rm eff} \lesssim 4500$\,K), low-gravity ($\log g \lesssim 0$) atmospheres with strong $\alpha$-element perturbations --- the kurucz-a1 emulator's prior is far enough from the true converged solution that \texttt{atlas\_py} can iterate into all-NaN before recovering. Defensive guards trap the failure with a clean error rather than silently writing a degenerate atmosphere; the recommended escape hatch is a \emph{neighbour warm-start}, in which a converged neighbour cell's atmosphere is used as the initial guess with the target's chemistry rewritten on top.

\paragraph{Performance.} The Python implementation is currently within a small factor of the Fortran original for equivalent calculations. Since the codebase is readable, well-structured Python with Numba-JIT inner loops, we expect that AI coding agents --- the same tools used to create this reimplementation --- can systematically profile and optimize the bottlenecks, making this a tractable engineering task rather than a fundamental limitation.

\section{AI usage disclosure}

A pure Python reimplementation of ATLAS12 and SYNTHE has been a long-standing aspiration in the stellar spectroscopy community --- discussed informally for decades --- but the software engineering effort has been prohibitive. The original Fortran is written in a pre-Fortran-77 style with fixed-format source, single-character variable names, pervasive use of computed GOTOs and EQUIVALENCE statements, and implicit typing throughout --- making it incompatible with modern \texttt{gfortran} without significant manual patching, and essentially unreadable without line-by-line consultation of Kurucz's own notes. The authors attempted this reimplementation repeatedly since the early days of large language models, but earlier models lacked the sustained context and code reasoning needed to navigate Kurucz's deeply interlinked subroutines. It was only with Anthropic's Claude Opus 4.6 (via Cursor) that the project became tractable --- though ``tractable'' still required extensive human guidance: several thousand US dollars in API tokens and several months of the authors' time directing, debugging, and validating the output. Every AI-generated routine was tested against the Fortran reference, first at the subroutine level and then end-to-end across a grid of atmosphere models. We view this as a demonstration of AI-assisted scientific software preservation: not fully autonomous, but a qualitative leap in what is feasible for a small team tackling a large legacy codebase.

\section{Acknowledgements}

This project is dedicated to the memory of Robert L.\ Kurucz (1944--2025), whose ATLAS and SYNTHE codes, atomic and molecular line lists, and freely shared data have served the astronomical community for over half a century.

\bibliographystyle{aasjournal}
\bibliography{paper}

\end{document}